\documentstyle[12pt]{article}
\newcommand{\swp}{\sum_{N=0}^{D-3}}
\newcommand{\SMI}{
                  \vspace{-13cm}
                  \begin{flushright}
                     SMI-06-97 \\
                  hep-th/9702079
                  \end{flushright}
                  \vspace{12cm}
                 }
\newcommand{\MIAN}{
                  \small \it Steklov Mathematical Institute,\\
\small\it Gubkin St. 8, GSP-1, 117966, Moscow, Russia,
                  }

\newcommand{\MIPT}{
                   \small
                   \it Department of General and Applied Physics,\\
                   \small
                   \it Moscow Institute of Physics and Technology\\
                   \small
                   \it Institutski per., 9, Dolgoprudnyi, Moscow region,
                   Russia
                  }
\font\twmsbm=msbm10 scaled 1200
\font\nmsbm=msbm9
\newfam\bbold
\textfont\bbold=\twmsbm
\scriptfont\bbold=\nmsbm
\scriptscriptfont\bbold=\nmsbm

\font\twscr=rsfs10 scaled 1200
\font\nscr=rsfs10
\newfam\script
\textfont\script=\twscr
\scriptfont\script=\nscr
\scriptscriptfont\script=\nscr
\def\Cal#1{{\fam\script#1}}
\author{I.~Ya.~Aref'eva\thanks{e-mail: arefeva@class.mi.ras.ru}
          \\ \MIAN
\\ M.~G.~Ivanov
              \\ \MIPT
\\
I.~V.~Volovich\thanks{e-mail: volovich@class.mi.ras.ru}
             \\ \MIAN}
\date{~}
\title{Non-extremal Intersecting p-branes\\
                in Various Dimensions }
\begin{document}
\maketitle
\SMI
\begin{abstract}

  Non-extremal intersecting p-brane solutions of gravity coupled
with several antisymmetric fields
and dilatons
in various space-time dimensions are constructed.
The construction uses the same algebraic method of finding solutions
as in the extremal case and a modified ``no-force'' conditions.
We justify the ``deformation'' prescription.
It is shown that the
non-extremal intersecting p-brane solutions
satisfy harmonic superposition rule and
the intersections of non-extremal p-branes are
specified by the same  characteristic equations for the incidence matrices
as for  the extremal p-branes. We show that $S$-duality
holds for non-extremal p-brane solutions. Generalized $T$-duality
takes place under additional restrictions to the
parameters of the theory,
which are the same as in the extremal case.

\end{abstract}
\newpage

 A recent study of duality \cite{HT}--\cite{Sen} and the microscopic
interpretation of the Bekenstein-Hawking entropy within
string theory \cite{Vafa,Hor} has stimulated  investigations of the
intersecting (composite) p-brane solutions.
An extension of the D-brane entropy counting to near-extremal black
p-branes is a problem of the valuable interest.
There has been recently considerable progress in the
study of classical extremal
\cite{TS} -\cite{AVVV}
and non-extremal p-brane solutions \cite{host}-\cite{Ohta}
in higher dimensional gravity coupled with matter.

Heuristic scheme of construction of p-brane intersections
was based on string theory representation of the branes,
duality and supersymmetry. This construction involves the
harmonic function superposition rule for the intersecting p-branes.
The rule was formulated in \cite{TS} based on \cite{ToPa1}
for extremal solutions in $D=11$ and $D=10$ space-time
dimensions. This rule was proved
in \cite{ArRy} in arbitrary dimensions by using an algebraic
method \cite{AV,AVV,IN} of solutions of the Einstein equations.

It has been shown that
there is a  prescription for ``deformation''
of a certain class of extremal $p$-branes
to give  non-extremal ones \cite{host,DLP}.
The harmonic function superposition rule for the intersecting p-branes
has also been extended to  non-extremal solutions with a single
``non-extremality'' parameter specifying a deviation from the
BPS-limit~\cite{CT}.
Non-extremal black hole solutions from intersecting M-branes
which are characterized by two non-extremal deformation parameters
have found in \cite{Ohta}.

 Another approach to the construction of p-brane solutions
was elaborated in the papers
\cite{AV,AVV,IN,IvMel,ArRy,AEH,AVVV}.
In these papers a general class of
p-brane intersection solutions was found. One starts from
the equations of motion and by using a
special ansatz for the metric and the matter
fields specified by the incidence matrix
one reduces them to  the Laplace equation and to
a system of algebraic equations.

The aim of this letter  is
to extend this approach to the non-extremal
case and to derive
an explicit formula (\ref{genmetr}), see below, for solutions.
We justify the
harmonic function superposition rule and show that
S- and T-dualities for intersecting non-extremal p-branes
hold.

 Let us consider the theory with the following action
 \begin{equation}\label{act}
   I=\frac{1}{2\kappa ^{2}}
      \int d^{D}X\sqrt{-g}
      \left(
         R-\frac{1}{2}(\nabla\vec\phi)^2-
         \sum_{I=1}^{k}
         \frac{e^{-\vec\alpha^{(I)}\vec\phi}}{2(d_I+1)!}
         F^{(I)2}_{d_I+1}
      \right).
 \end{equation}
 where $F^{(I)}_{d_I+1}$ is a $d_I+1$ differential form,
 $F^{(I)}_{d_I+1}=dA_{d_I}$, $\vec\phi$ is a  set of
 dilaton fields.

 Extremal solutions have been found starting from the
 following ansatz for the metric
 \begin{equation}\label{3}
  ds^2=\sum_{i,k=0}^{D-s-3}e^{2F_i(x)}\eta_{ik}dX^i dX^i
      +e^{2B(x)}\sum_{\gamma=D-s-2}^{D-1}dx^\gamma dx^\gamma,
 \end{equation}
 where $\eta_{\mu\nu}$ is a flat Minkowski metric.

 In the non-extremal case we shall start from the
 following ansatz for metric. We choose a direction $i_{0}$
 belonging to $0<i<D-s-3$ and consider the following ansatz

 \begin{equation}\label{Dmetr}
   ds^2=\sum_{i,k=0}^{D-s-3}e^{2F_{i}(r)}
          \eta_{ik}f^{\delta_{ii_0}}(r)dX^i dX^k
         +e^{2B(r)}
         \left(
           r^2 d\Omega^2_{s+1}+f^{-1}(r)dr^2
         \right),
\end{equation}
Here $r=\sqrt{x^{\gamma}x^{\gamma}}$ and
$f(r)$ is an arbitrary function of $r$.
We shall work in the following gauge
 \begin{equation}\label{gauge}
    \sum_{L=0}^{D-3}F_L=0,
 \end{equation}
where we assume $F_L=B$ for every $L=D-s-2,\dots,D-1$.
In the extremal case such gauge is Fock--De~Donder
one, but in non-extremal case it is not true.

To specify the ansatz for the antisymmetric fields
we will use incidence matrices \cite{ArRy,AIR}
 \begin{eqnarray}
                         \label{4e}
  \Delta^{(I)}&=&(\Delta ^{(I)}_{aL}), ~~a=1,\dots,E, ~~L=0,\dots,D-1,\\
                         \label{4m}
  \Lambda^{(I)}&=&(\Lambda ^{(I)}_{bL}), ~~b=1,\dots,M, ~~L=0,\dots,D-1.
 \end{eqnarray}

 The entries of the incidence matrix are equal to 1 or 0.
 Their rows correspond to independent branches of  the electric
 (magnetic) gauge field and  columns refer to the
 space-time indices of the metric (\ref{3}).
 We assume
 $\Delta_{a0}=1,\;\Delta_{a\alpha}=0,
 \;\Lambda_{b0}=0,\;\Lambda_{b\alpha}=1$ and

 \begin{equation}
   F^{(I)}=\sum_{a=1}^{E^{(I)}} dA^{(i)}_a
          +\sum_{b=1}^{E^{(I)} } F^{(I)}_b,
 \end{equation}
 where the coefficients of the differential
 forms $A^{(I)}_a$ and $F^{(I)}_u$ are
 \begin{eqnarray}
 \label{ea}
   A^{(I)}_{a~M_1\cdots M_{d_I}}&=&
     \epsilon_{M_1\cdots M_{d_{\Lambda(a)}}}h_a e^{D_a}
     \prod_{i=1}^{d_I}\Delta_{aM_i},\\
 \label{ma}
   F_b^{(I)~M_0\cdots M_{d_I}}&=&
     \epsilon^{M_0\cdots M_{d_I} \alpha}
     \frac{h_b}{\sqrt{-g}}e^{\alpha^{(I)}\phi}
     \partial_{\alpha}  e^{D_b}
     \prod_{i=0}^{d_I}\Lambda_{bM_i}.
 \end{eqnarray}
 Here $\epsilon^{01\cdots}=
      \epsilon_{01\cdots}=1$
 are totally antisymmetric symbols,
 $D_a$ and $D_b$ are functions of $X^\alpha$.
The product $\prod_{i=1}^{d_I}\Delta_{aM_i}$
selects the non-zero components of electric part the form $A$  and
the product $\prod_{i=0}^{d_I}\Lambda_{bM_i}$
selects the non-zero components of
magnetic components of its strength.

 We will use Einstein equations in the form $R_{KL}=G_{KL}$,
 there $G_{KL}$ is related to the stress-energy tensor $T_{KL}$ as
 \begin{equation}
   G_{KL}=T_{KL}-\frac{g_{KL}}{D-2}T_P^{~~P}.
 \end{equation}

 For the above  ansatz  the tensor $G_{KL}$ is

 \begin{eqnarray}\label{tensorG}
  G_{KL}&=&\frac{1}{2}\partial_K\phi\partial_L\phi\\
          &+&\sum_R\frac{h_R^2}{2}
            e^{2\tilde F_L-2B+\tilde{{\cal F}}_R}
            \left(
              -\partial_K D_R\partial_L D_R
              -\varsigma_R\eta_{KL}
              \left\{
                \Delta_{RL}-\frac{d_R}{D-2}
              \right\}
              (\partial D_R)^2
            \right),\nonumber
 \end{eqnarray}
 where $R=a$ or $b$, $\Delta_{bL}=\Lambda_{bL}$,
 \begin{eqnarray}
   \tilde{{\cal F}}_R&=&2D_R
             -2\sum_{N=0}^{D-3}
             \Delta_{RN}F_N-\alpha\phi,
 \end{eqnarray}

\begin{equation}
\varsigma_a=-1, ~~\varsigma_b=+1,
\end{equation}
and
\begin{equation}
 \tilde F_L~~\mbox{is}~~ F_L ~~\mbox{if}~~ L\not=i_0,r,~~
\tilde F_{i_0}=F_{i_0}+1/2\ln f,~~
\tilde F_r=F_r-1/2\ln f.
\end{equation}
Now let us suppose the following ``no-force''
condition
 \begin{eqnarray}\label{no-f}
   {\cal F}_R&\equiv&2C_R
             -2\sum_{N=0}^{D-3}
             \Delta_{RN}F_N-\alpha\phi=0
 \end{eqnarray}
where $C_{R}$ is an one-center function
\begin{equation}
                                           \label{H}
e^{-C_R}\equiv H_R=1+\frac{Q_{R}}{r^{s}},
\end{equation}
and $Q_R$ is a constant.

Under ``no-force'' conditions (\ref{no-f}) the field equations
for the electric components and the Bianchi identities
for magnetic components are reduced to
 \begin{equation}
  \Box D_R+(\partial D_R,\partial (D_R-2C_R))=0.
 \end{equation}

This equation for the one-center functions $C_R$
has the following solution
\begin{equation}
                                    \label{tH}
 e^{-D_R}\equiv\tilde H_R=
 1+\frac{Q_{R}+{\Cal F}_{R}}{r^{s}-{\Cal F}_{R}},
\end{equation}
where ${\Cal F}_R$ is a constant.

Under the ``no-force'' conditions the $G$-tensor
can be written in abridged notations as
 \begin{eqnarray}
  G_{KL}&=&g_{KL}\frac{\varsigma_R h^2_R}{2\kappa^2}
       \left[\hat \Delta_{RK}-\frac{d_R}{D-2}\right]
       e^{2(D_R+{\cal F}_R-C_R-B)}(\partial D_R)^2,\\
  \hat \Delta_{RM}&=&\left\{\begin{array}{ll}
                \Delta_{RK},&K\not=D-1\\
                \Delta_{RK}-\varsigma_R,&K=D-1
                \end{array}\right. .
 \end{eqnarray}

Field equation for dilaton reads
 \begin{equation}
   \label{fi}
   f\Box\phi+(\partial\phi,\partial f)=
   -\sum_R\varsigma_R\alpha_R
   \frac{h_R^2}{2} e^{2(D_R-C_R)} (\partial D_R)^2.
 \end{equation}


 To solve the Einstein equations let us write
the Ricci tensor for the metric (\ref{Dmetr}) explicitly.
For simplicity  we calculate it in the stereographic
parametrization of sphere,
\begin{equation}
   d\Omega^2_{s+1}=\frac{4 dz^{\hat\alpha}dz^{\hat\alpha}}
                  {(1+z^{\hat\gamma}z^{\hat\gamma})^2},
 \end{equation}
 at the point $z^{\hat\alpha}=0$. We get

 \begin{equation}
   R_{ij}=-g_{ij}
            e^{-2B}[f\Box F_{i}+
            (\partial F_{i},\partial f)],
 \end{equation}

 \begin{equation}\label{i0-i0}
   R_{i_0i_0}=-g_{i_0i_0}
            e^{-2B}\left[f\Box F_{i_0}+
            (\partial F_{i_0},\partial f)+
         \frac{\triangle f}{2f}\right],
 \end{equation}

 \begin{equation}\label{hat-hat}
   R_{\hat\alpha\hat\beta}=-g_{\hat\alpha\hat\beta}
            e^{-2B}[f\Box F_{\hat\alpha}+
            (\partial F_{\hat\alpha},\partial f)]
          +4\eta_{\hat\alpha\hat\beta}[s(1-f)-r\partial_r f],
\end{equation}

 \begin{equation}\label{r-r}
   R_{rr}=-\swp \partial_r A^2_{N}-\eta_{rr}f^{-1}
            [f\Box B+(\partial F_{i_0},\partial f)]
            -\frac{\triangle f}{2f}.
\end{equation}

 To kill the last terms in (\ref{i0-i0})--(\ref{r-r})
 we suppose that
 \begin{equation}\label{fCon}
   s(1-f)-r\partial_r f=0.
 \end{equation}
 Then
 \begin{equation}
   f=1-\frac{2\mu}{r^s},
 \end{equation}
 where $\mu$ is a constant.

Under condition (\ref{fCon}) the Ricci tensor takes the form

 \begin{equation}
   R_{\lambda\mu}=-g_{\lambda\nu}
            e^{-2B}[f\Box F_{\lambda}+
            (\partial F_{\lambda},\partial f)],
 \end{equation}
 \begin{equation}
   R_{rr}=-\swp \partial_r A^2_{N}-\eta_{rr}f^{-1}
            [f\Box B+(\partial F_{i_0},\partial f)].
\end{equation}

 Combining the Einstein equations and the field equation
 for dilaton we see that it is natural to set
 \begin{equation}
    \label{mN}
    f\Box C_R+(\partial C_R,\partial f)-
    e^{2(D_R-C_R)}(\partial D_R)^2=0.
 \end{equation}
 Using the ``no-force'' conditions
 we derive equations for the constants $h_R$ and
 the characteristic equations,
 which are the same as in the extremal case:
 \begin{equation}\label{char}
   (1-\delta_{RR'})
   \left\{
     {\vec\alpha_R\vec\alpha_{R'}\over 2}
     -{d_R d_{R'}\over D-2}
     +\sum_{L=0}^{D-3}\Delta_{RL}\Delta_{R'L}
   \right\}=0,
 \end{equation}

 Substituting $C_R$ and $D_R$ into equation (\ref{mN})
 we get
 \begin{equation}
   \Cal{F}_R=-Q_R\pm\sqrt{Q_R^2+2\mu Q_R}.
 \end{equation}
 As in the case non-extremal M-branes \cite{CT} it is
 convenient to parametrize the deformations of the
 harmonic functions $H_R$ in the following way
 \begin{eqnarray}
 \label{H2H}
   H_R&\rightarrow&\tilde H_R,\\
   \tilde H_R&=&1+
          \frac{Q_R+\Cal{F}_R}
          {r^s-\Cal{F}_R},\\
 \label{eQ}
   \Cal{Q}_R&=&2\mu\sinh\gamma_R\cosh\gamma_R,\\
   Q_R&=&2\mu\sinh^2 \gamma_R,\\
 \label{eF}
   \Cal{F}_R&=&\Cal{Q}_R-Q_R.
 \end{eqnarray}

 From Einstein equation for $(rr)$-component one
 can find the following restriction for $i_0$
 \begin{equation}\label{i0R}
   \Delta_{ai_0}=1,\qquad \Lambda_{bi_0}=0.
 \end{equation}

Let us write the final expression for the metric

  $$
   ds^2=\prod_{I=1}^k
        \left(
          H_{1}^{(I)} H_{2}^{(I)}\cdots
          H_{E_I}^{(I)}
        \right)^{2u^{(I)}
        \sigma^{(I)}}
        \left(
          H_1^{(I)} H_2^{(I)}\cdots H_{M_I}^{(I)}
        \right)^{2t^{(I)}
        \sigma^{(I)}}
 $$
  \begin{equation}\label{genmetr}
   \left\{
     \sum_{L=0}^{D-s-3}
     \prod_{I=1}^k
     \left(
       \prod _{a}{H_a^{(I)}}^{\Delta_{ai}^{(I)}}
       \prod _{b}{H_b^{(I)}}^{1-\Lambda_{bi}^{(I)}}
     \right)^{-\sigma^{(I)}}f^{\delta _{i_{0}K}}\eta_{KL}
     dy_K dy_L
     +d\Omega_{s+1}^2+f^{-1}dr^2
   \right\},
 \end{equation}
 where
 \begin{equation}\label{t-u}
   t^{(I)}=\frac{D-2-d_I}{2(D-2)},\qquad
   u^{(I)}=\frac{    d_I}{2(D-2)}.
 \end{equation}
 Constants in the ansatz for the antisymmetric fields
 have the form
 \begin{equation}
   \left.h_a^{(I)}\right.^2=
   \left.h_b^{(I)}\right.^2=
   \sigma^{(I)},\qquad
   \mbox{where}\qquad
   \sigma^{(I)}=
   \frac{1}{t^{(I)}d_I+\left.\vec\alpha^{(I)}\right.^2/4}.
 \end{equation}
The incident matrices satisfy to the
characteristic equations (\ref{char}).

 Let us note that the harmonic superposition rule is obvious from the
 expression (\ref{genmetr}).  Since the characteristic equations
 for the non-extremal case are the same as for the extremal case the
 duality properties are the same in both cases. In particular,
 $S$-duality takes place for all values of parameters, as to
 $T$-duality it takes place under the same restrictions on
 the parameters of the theory. Note that one can perform the $T$-duality
 transformation only along $i$-directions such that $i\neq i_{0}$.

 All above results may be generalized for the space-time
 with an arbitrary signature. Kaluza-Klein theory with
 extra time-like dimensions has been considered
 in \cite{AV85}.
 One deals with a modification of
 the action (\ref{act}) in which the standard ``$-$'' signs before
 the $F^2$ terms are changed by $-s_I$, where $s_I=\pm 1$, and
 $\sqrt{-g}$ is changed by $\sqrt{|g|}$.

 In this case one has to replace in the metric (\ref{genmetr})
 the term $f^{-1}dr^2$ by the $\eta_{rr}f^{-1}dr^2$, and
 set the following new equations
 \begin{eqnarray}
   r&=&\sqrt{|\eta_{\alpha\beta}x^\alpha x^\beta|}
     = \sqrt{\eta_{rr}\eta_{\alpha\beta}x^\alpha x^\beta},\\
   d\Omega^2_{s+1}&=&\frac{4\eta_{\hat\alpha\hat\beta}
                   dz^{\hat\alpha}dz^{\hat\beta}}
                   {(1+\eta_{rr}\eta_{\hat\gamma\hat\delta}
                   z^{\hat\gamma}z^{\hat\delta})^2},\\
   \eta_{\hat\alpha\hat\beta}&=&{\rm diag}(\pm1,\dots,\pm1),
 \end{eqnarray}
 where the matrix
 $$
   \left(\begin{array}{c|c}
        \eta_{rr}& 0\\
        \hline
        0 & \eta_{\hat\alpha\hat\beta} \\
   \end{array}\right),
 $$
 has the same signature as $\eta_{\alpha\beta}$.

 For the incidence matrix instead of two old restrictions
 \begin{equation}\label{oldR}
   \Delta_{a0}=1,\qquad
   \Lambda_{b0}=0,
 \end{equation}
 one has
 \begin{equation}\label{newR}
   s_I\prod\limits_{L=0}^{D-1}(\eta_{LL})^{\Delta_{RL}}=\varsigma_R.
 \end{equation}
 So one can see that in the case of standard signature
 conditions (\ref{oldR}) always guarantee the existence
 of $i_0$ ($i_0=0$ always exist), but in the general case
 the conditions (\ref{i0R}) are not trivial.


\vspace{1cm}

 To summarize, using an algebraic method of solution
 of Einstein equations in various dimensions
 we  have  constructed the non-extremal
 intersecting p-brane solutions (\ref{genmetr})
 which satisfy the
 harmonic function superposition rule and possess $S$- and
 $T$-dualities. The intersections of non-extremal p-branes
 are controlled by the same characteristic equations as
 for the extremal cases.


\section*{Acknowledgments}
This work is partially supported by the RFFI grants 96-01-00608
 (I.A.) and 96-01-00312 (M.I. and I.V.).


\end{document}